\documentstyle[aps,twocolumn,prb,epsf]{revtex}

\begin{document}
\draft

\twocolumn[\hsize\textwidth\columnwidth\hsize\csname@twocolumnfalse\endcsname
\title{Optical spectral weights and the
ferromagnetic transition temperature of CMR manganites: relevance of
double-exchange to real materials}  \author{A.\ Chattopadhyay$^1$, A.\
J.\ Millis$^2$ and  S.\ Das Sarma$^1$} 
\address{ 
         $^1$ Department of Physics, University of Maryland\\
              College Park, MD 20742\\
         $^2$ Center for Materials Theory\\
              Department of Physics and Astronomy, Rutgers University\\       
              Piscataway, NJ 08854\\ }
\date{\today}
\maketitle

\widetext
\begin{abstract}
\noindent

We present a thorough and quantitative comparison of double-exchange 
models to experimental data on the colossal magnetoresistance manganese
perovskites. Our results settle a controversy by showing that physics 
beyond double-exchange is important even in La$_{0.7}$Sr$_{0.3}$MnO$_3$, 
which has been regarded as a conventional double-exchange system. 
We show that the crucial quantity for comparisons of different calculations
to each other and to data is the conduction band kinetic energy $K$, which is
insensitive to the details of the band structure and can  be experimentally
determined from optical conductivity measurements. The seemingly complicated
dependence of $T_c$ on the Hund's coupling $J$ and carrier concentration $n$ 
is shown to reflect the variation of $K$ with $J$, $n$ and
temperature. We present results for the optical conductivity which  allow
interpretation of experiments and show that a feature previously  interpreted
in terms of the Hund's coupling was misidentified. We also
correct minor errors in the phase diagram presented in previous work.

\end{abstract}

\vspace*{1cm}]

\narrowtext

\section{INTRODUCTION}

The colossal magnetoresistance (CMR) rare earth manganese perovskites
first attracted attention in the 1950s because of the range of magnetic 
and structural transitions they display. Recent interest has been 
revived by the extremely large ("colossal") magnetoresistance displayed 
by some CMR materials, coupled with their rich phase diagram\cite{tokurabook}. 
Despite this long and intense study, much of the physics of CMR remains
controversial, with basic issues still subject to debate. In this paper
we address two such issues: the first is in what sense the standard
'double-exchange only' model (defined below) describes the physics in the
regions of the phase diagram where the ground state is a ferromagnetic metal.
The second is the interpretation of the optical conductivity spectrum, and in
particular which portions of the observed spectrum pertain to the conduction
band electrons responsible for the interesting physics of the CMR materials. In
addition to its direct relevance to CMR, we believe our work is of broader
significance for the theory of interacting electrons in solids,
as a contribution to the fundamental issues of the quantitative comparison of
many-body calculations to the properties of materials, and to the
interpretation of the optical spectra of correlated electron materials.

It is generally agreed that a crucial aspect of CMR physics is
'double-exchange'\cite{Degennes60}: the mobile carriers (which, for CMR
materials, are $Mn$ $e_g$ symmetry  $d$-electrons) are strongly coupled
ferromagnetically to localized core spins ($Mn$ $t_{2g}$ symmetry electrons).
This means that core spin alignment dictates carrier motion, which in general
leads to ferromagnetism. What, if any, additional physics beyond
double-exchange is important for the manganites is still
debated\cite{tokurabook}. For example, some authors have argued that
double-exchange only models correctly predict the magnetic transition
temperatures of CMR materials\cite{Furukawa,Calderon98}, while others have
argued that they do not\cite{DagottoScience,Millis95}. Published calculations
of $T_c$\cite{Furukawa,Yunoki98} have not resolved the issue because the 
results depend on model parameters such as bandwidth, interaction  strength and
carrier density, often chosen arbitrarily or in a manner inconsistent with
the physics of the manganites. We
attempt to settle the question by establishing the full phase diagram and by 
presenting precise calculations of $T_c$
within a specific model. Most importantly, by showing how the parameters
on which the calculated  $T_c$ depends can be related to measured properties
of real materials and to parameters used in other calculations. 

Another important set of issues concerns the optical 
conductivity, which in the CMR materials has a strong dependence on 
chemical composition, frequency, temperature and magnetic field
\cite{tokurabook,Okimoto95,Okimoto97,Quijada98,Noh98,Noh99}.
It is widely believed that information extracted from these data will be
helpful in elucidating the physics of CMR, but this goal has not yet been 
fully achieved. Even the basic question of which parts of the observed
spectrum are due to the conduction band degrees of freedom is still
not settled\cite{Okimoto95,Okimoto97,Quijada98}. We show here that a 
comparison of the data to
the theoretically calculated magnitude and temperature dependence of optical
spectral weight, and of its relationship to $T_c$ can resolve this issue. Our
results  indicate that
several previous analyses\cite{Okimoto95,Okimoto97} underestimate the actual
conduction band spectral weight, and suggest that a previous paper by several
authors including one of us\cite{Quijada98} misidentified a key feature in the
data. This misidentification arose primarily from reliance on a
theoretical calculation based on inappropriate system parameters.

In this paper we perform a systematic and quantitative comparison of the
predictions of double-exchange-only models to data. We incorporate "real
materials" aspects via a tight-binding fit to the band structure; in 
these materials a simple nearest neighbour hopping model represents the 
real material parameters adequately. In particular, we show that the 
crucial quantity for comparison of calculations to data or to
other calculations is the conduction-band kinetic energy, a local expectation
value related to the electron hopping amplitude and defined more precisely
in Eq.~(\ref{kin1}). Because this kinetic energy is a local
quantity, it can be easily and reliably calculated and is insensitive to
the fine details of the band structure. When expressed in terms of 
the kinetic energy, the apparently disparate results of a variety of
computations are seen to be in quantitative agreement. More importantly,
because the kinetic energy may be determined from optical conductivity
experiments\cite{Millis90,Baeriswyl86}, our results enable a detailed
comparison of theory to experiment. 

To treat the many-body physics of double-exchange, we use the highly
successful dynamical mean-field method\cite{georges96}. This provides a
detailed and apparently reliable treatment of the local correlations which are
crucial to the conclusions we wish to draw. We will argue that the effect of
the omitted physics, which has to do with intersite fluctuations, may be
reliably estimated for the issues of interest to us and is not too large.

The application of the dynamical mean-field method to double-exchange
systems was pioneered by Furukawa\cite{Furukawa}, whose important work 
established the basic usefulness of this method and presented many 
valuable results concerning the phase diagram, conductivity and other 
properties. However, his work was incomplete in some respects, incorrect 
in some details, and in many cases employed physically irrelevant parameters. 
The full treatment we present here is therefore needed.

We believe that the analysis presented in this paper may
be useful in the general context of the development of a predictive theory of
dynamical and ordering phenomena in systems, such as transition metal oxides,
with strong interactions. This is an important 
goal of electronic condensed matter physics; and recent advances in 
computational power and in the techniques of many-body physics suggest
that it may be obtainable. 
The dynamical mean field method\cite{georges96} seems particularly
promising in this respect. It is
computationally tractable, includes incoherent and inelastic effects, and can
be combined with conventional band theory. Investigations of the
extent to which this method is useful in the computation of real materials
properties are urgently needed. The work described below is such an
investigation. In this context we point out that one of the great 
advances in the theory of equilibrium critical phenomena arose from a 
careful and detailed application of mean-field theories. We 
believe that a similar opportunity may be present in the application 
of the dynamical mean field theory to strong correlation physics. 

The rest of the paper is organized as follows. Section II defines the
double-exchange-only model, and explains how its parameters should be defined
and related to experiments. Section III presents the dynamical mean-field
formalism. Section IV presents the numerical and analytical results for the
phase  diagram, and corrects what seems to be a minor error in the phase
diagram proposed in \cite{Yunoki98}. Section V presents results for the
spectral weight (integrated area) in different regions of the optical
conductivity spectrum. Section VI provides a summary of our theoretical 
results and a detailed interpretation of experimental data, particularly 
of optical conductivity and its relation to $T_c$. Readers interested only
in this aspect may proceed directly to this section. Section VII is a
conclusion and presents some possible directions for future research. 

\section{MODEL}

The double-exchange-only (DE) model involves electrons (with orbital 
indices $a,b$) moving in a band structure defined by a hopping
matrix $t_{ij}^{ab}$ and a chemical potential $\mu$, and connected by a Hund's
coupling ($J > 0$) to core spins $\bf{S}$, which we take to be classical.
We denote the  operator creating an electron of spin $\alpha$ on orbital $a$ of
site $i$ by $d_{ia\alpha}^{\dagger}$ and define the double-exchange only
Hamiltonian $H_{DE}$ by\cite{Kubo72}
\begin{eqnarray}
H_{DE} = & - & \sum_{<ij>,ab} t_{ij}^{ab} d_{ia\alpha}^{\dagger} d_{jb\alpha} - 
\mu \sum_{ia \alpha} d_{ia\alpha}^{\dagger} d_{ia\alpha} \nonumber \\ 
 & - & J \sum_{ia\alpha\beta} {\vec S}_{i} \cdot d_{ia\alpha}^{\dagger}
{\vec \sigma}_{\alpha\beta} d_{ia\beta}
\,.
\label{Hdex}
\end{eqnarray} 
Here $t_{ij}^{ab}$ is the amplitude to hop from site $i$, orbital $a$ to 
site $j$, orbital $b$. The calculated band structure\cite{Pickett96}
is well fit by a  $t_{ij}^{ab}$ which involves only nearest-neighbour hopping. 

A crucial quantity is the electron kinetic energy
$K$, defined by
\begin{eqnarray}
K & = & \frac{-2}{Z N_{sites}} \sum_{<ij>}\left<t_{ij}^{ab}
d^{\dagger}_{ia\sigma}d_{jb\sigma}\right> \nonumber \\
& = & \frac{2}{Z} \int
\,\frac{d^{d}p}{(2 \pi)^d} 
\epsilon^{ab}_{p}\left<d^{\dagger}_{pa\alpha}d_{pb\alpha}\right> \,,
\label{kin1}
\end{eqnarray}
where $\epsilon^{ab}_{p}$ is the dispersion obtained by Fourier 
transforming $t_{ij}^{ab}$, $Z$ is the number of nearest neighbours  and 
$d$ is the spatial dimension. In the limit $J\rightarrow \infty$, $K$ is
the only relevant energy scale in the model and in particular the magnetic
transition temperature $T_c$ has been found to depend only on 
$K$ \cite{Degennes60,Kubo72,Millis96b}. 
$K$ is also a suitable quantity for comparison to experiment
because, in the physically realistic limit of nearest-neighbour 
hopping, $K$ may be determined from an analysis of the optical 
conductivity \cite{Quijada98,Millis90}. We would like to mention that this 
$K$ includes a sum over all bond directions and is 3 times larger than the
$K$ defined in Quijada et al.

\section{FORMALISM}

In this paper we study the magnetic transition temperature
and kinetic energy of $H_{DE}$ as functions of carrier density
$n$ and Hund's coupling using 
the dynamical mean field approximation, supplemented by an analytical 
treatment of the $J \rightarrow 0$ limit and by comparison to results of Monte
Carlo \cite{Yunoki98} and series expansions\cite{Roder97}.

The dynamical mean field method has been discussed extensively 
elsewhere \cite{georges96} and therefore we omit formal details. This
scheme was first obtained in the limit $d \rightarrow \infty$ with $t_{ij}\sim
1/\sqrt{d}$, and is often referred to as the infinite dimensional limit, but
the central approximation is actually the neglect of the momentum 
dependence of the electron self-energy, $\Sigma$. While this becomes 
exact in the limit $d \rightarrow \infty$, it is quite a good
approximation in $d=3$\cite{georges96}, where the momentum dependence of
$\Sigma$ is quite weak.

To make the importance of the neglect of momentum dependence clear and to
establish notation we sketch a derivation of the dynamical mean field
equations. A momentum independent $\Sigma$ implies that all of 
the many-body physics may be derived from the local  (momentum-integrated)
Green's function ${\bf G}_{loc}$, defined by 
\begin{equation}
{\bf G}_{loc}(\omega) = Tr \int\, \frac{d^{d}p}{(2 \pi)^d} 
{\bf{G}}(p,\omega)
\label{gloc1}
\end{equation}
where the Green's function, a matrix in orbital ($ab$) and spin
($\alpha \beta$) space, 
is
\begin{equation}
{\bf{G}}(p,\omega) =
\left[\omega +\mu-\epsilon_{p}^{ab} -
\Sigma_{\alpha \beta}^{ab}(\omega)\right]^{-1}
\,.
\label{green}
\end{equation}
It is convenient to define the density of states summed over bands 
$N(\epsilon) = \sum_{\lambda} \int \,\frac{d^{d}p}{(2 \pi)^d}
\delta(\epsilon - \epsilon_{p}^{\lambda})$, where $\epsilon_{p}^{\lambda}$
are the eigenvalues of $\epsilon^{ab}_{p}$. We shall be interested in the 
pseudo-cubic manganese perovskites in the doping regimes where orbital
order is not important. For these materials
$N_{\lambda}(\epsilon) = \int \,\frac{d^{d}p}{(2 \pi)^d}
\delta(\epsilon - \epsilon_{p}^{\lambda})$ is independent of $\lambda$; 
thus we use $N(\epsilon)= n_{orb} N_{\lambda}(\epsilon)$, where we have
introduced the number of orbitals $n_{orb}$. For the physically relevant
case, $n_{orb}=2$, but the case $n_{orb} =1$ has been considered by other
workers\cite{Furukawa}, so we retain $n_{orb}$ as a variable to  facilitate
comparisons. The local Green's function is then proportional to the unit
matrix in $ab$ space, ${\bf G} = G_{loc,\alpha \beta} \delta_{ab}$
and the coefficient is
\begin{equation}
G_{loc,\alpha \beta}(\omega) = \int\, d\epsilon N(\epsilon)
\frac{1}{\omega +\mu-\epsilon-\Sigma_{\alpha \beta}(\omega)}
\,.
\label{gloc2}
\end{equation}
$G_{loc}$ depends only on frequency and is therefore the solution 
of a single-site problem. This problem is specified by a mean-field function
$b_{\sigma}(\omega)$ which, in the model of present interest, is related
to $G_{loc}$ via\cite{Furukawa,Millis96b}
\begin{equation}
G_{loc,\alpha \beta}(\omega) = \int \,d^{2}{\vec{\Omega}}
\frac{P(\vec{\Omega})} {b_{\alpha \beta}(\omega)-
J{\vec{S}}\cdot{\vec{\sigma}_{\alpha \beta}}};  {\vec{\Omega}} =
{\vec{S}}/\left|{\vec{S}}\right| 
\label{gloc3}
\end{equation}
with 
\begin{equation}
P({\vec{\Omega}}) = \frac{1}{Z_{loc}} 
\exp\left[\sum_{n} Tr \ln \left[b_{\alpha \beta}(i\omega_{n})- 
J{\vec{S}}\cdot{\vec{\sigma}_{\alpha \beta}}\right]\right]
\,,
\end{equation}
and $Z_{loc}$ ensures that $\int \,d^{2}{\vec{\Omega}}P({\vec\Omega})=1$. 
The self-energy is defined by 
\begin{equation}
\Sigma_{\alpha \beta}(\omega)=b_{\alpha \beta}(\omega)-
G_{loc,\alpha \beta}^{-1}(\omega)
\label{self}
\end{equation}
and the mean-field function $b_{\alpha \beta}(\omega)$ is fixed by the
requirement that  Eq.~(\ref{gloc1}) holds with $G_{\alpha \beta}(p,\omega)$
defined by Eq.~(\ref{green}), $\Sigma$ defined by Eq.~(\ref{self}) and 
$G_{loc}$ by Eq.~(\ref{gloc3}). Using the momentum independence of the 
self-energies within this mean-field theory, the electron kinetic energy
(Eq.~(\ref{kin1})) can be written as  
\begin{eqnarray}
K  & \equiv & \frac{2}{Z} T \sum_{n} \int\,\frac{d^{d}p}{(2 \pi)^d}
\epsilon_{p} Tr [{\bf G}(p,i\omega_{n})] \nonumber \\
& = & \frac{2}{Z} T \sum_{n} Tr \left[G_{loc}(i\omega_{n})\right]^2
\label{kin2}
\,.
\end{eqnarray}

In a ferromagnetic state with magnetization direction $\hat{m}$ we 
have
\begin{equation}
b_{\alpha \beta}(\omega) = b_{0}(\omega) + b_{1}(\omega)\hat{m}\cdot
\vec{\sigma}_{\alpha \beta}
\label{b0b1}
\end{equation}
($b_{1}(\omega)=0$ in the paramagnetic state). If the spin axis is
chosen parallel to $\hat{m}$ then $b_{\alpha \beta}$ becomes a 
diagonal matrix with components parallel 
($b_{\uparrow} =  b_{0}+b_{1}$) and antiparallel 
($b_{\downarrow} =  b_{0}-b_{1}$) to $\hat{m}$.

The precise form of the equation for $b_{\sigma}$ depends on the
form of the density of states $N(\epsilon)$, but the important
behaviour of the physical observables do not, as long as $N(\epsilon)$ 
has a finite second moment
($\int\, d\epsilon N(\epsilon)\epsilon^{2} < \infty$). We perform
our calculations on the Bethe lattice with a semi-circular 
density of states per orbital
\begin{equation}
N(\epsilon) = \frac{\sqrt{4 t^{2} - \epsilon^2}}{2 \pi t^2}
\label{dos}
\,.
\end{equation}
Here $t$ is a bandwidth parameter; the full bandwidth of the non-
interacting ($J=0$) problem is $4t$. This corresponds to number of 
nearest neighbours $Z=2d$, as $Z\rightarrow \infty$ with $tZ$ fixed.
For this
$N(\epsilon)$, the self-consistent equations for the $b_{\sigma}$ are
\begin{eqnarray}
b_{\uparrow} & = & i \omega_{n}+ \mu+ \int^1_{-1} \,d(cos\theta)  
P(\theta) \times \nonumber \\
 &  & \frac{(-b_{\downarrow} + J cos \theta)}
{b_{\uparrow} b_{\downarrow} - (b_{\uparrow}-b_{\downarrow})
J cos \theta -J^{2}} \nonumber \\
b_{\downarrow} & = & i \omega_{n}+ \mu- \int^1_{-1} \,d(cos \theta) 
P(\theta) \times \nonumber \\
 &  & \frac{(b_{\uparrow} + J cos \theta)}
{b_{\uparrow} b_{\downarrow} - (b_{\uparrow}-b_{\downarrow})
J cos \theta -J^{2}} 
\label{mfeq}
\end{eqnarray}
and the local Green's function (Eq.~\ref{gloc3}) can be written in terms
of the mean-fields as 
\begin{equation}
G_{loc}^{\sigma}(i\omega_{n}) =  i\omega_{n}+\mu -b_{\sigma}
(i\omega_{n})
\,.
\end{equation}
The self energies are evaluated using Eq.~(\ref{self}) and the
full Green's function from Eq.~(\ref{green}).

We will also be interested in the optical conductivity, $\sigma$. The 
contributions to $\sigma$ from the states described by $H_{DE}$ are
obtained by couplng an electromagnetic field to $H_{DE}$ (Eq.~\ref{Hdex}) via
the Peierls coupling $t_{ij} \rightarrow t_{ij}e^{i A (r_{i}-r_{j})}$ and 
performing linear response. 
If the self-energy is momentum independent then there are no vertex
corrections in $\sigma$\cite{Mahan} and 
\begin{eqnarray}
\sigma_{ij}(i\Omega_{n}) & = & \frac{e^2}{i\Omega_{n}}
\Bigg[ S(\infty) - \sum_{\sigma} \int\,\frac{d^{d}p}{(2 \pi)^d}
\nonumber \\ 
T \sum_{m} & Tr &  \Big[{\bf  \gamma}_{p}^{i} {\bf G}_{\sigma}(p,i\omega_{m}) 
{\bf \gamma}_{p}^{j} {\bf G}_{\sigma}(p,i\Omega_{n}+i\omega_{m})\Big]
\Bigg]  \label{opt-cond1}
\end{eqnarray}
where ${\bf \gamma}_{p}$ is the current operator 
determined by the Peierls 
substitution; it includes both intra and inter-band transitions within the 
manifold of conduction band states. Since we concentrate on materials with
cubic symmetry, $\sigma_{ij} \sim \delta_{ij}$ and hereafter we suppress
the spatial indices on $\sigma$.

This conductivity obeys the sum rule\cite{Millis90,Baeriswyl86}
\begin{eqnarray}
S(\infty) & = & \frac{a^{d-2}}{e^2}\int_{0}^{\infty} \frac{2}{\pi} 
\,{d\Omega}\sigma(\Omega) \nonumber \\
& = & \sum_{i\delta ab \alpha} t^{ab}_{\bf{\delta}} {\delta^2}
\left<d^{\dagger}_{ia\alpha}d_{i+{\bf{\delta}}b\alpha} + H.c.
\right>
\label{spec-wt}
\,.
\end{eqnarray}
If the nearest neighbour hopping is dominant then 
$\int_{0}^{\infty} \frac{2}{\pi} \,{d\Omega}\sigma(\Omega)=
\frac{e^2}{a^{d-2}}K$. The experimentally measured optical conductivity 
includes additional transitions not described by $H_{DE}$ and obeys the
familiar f-sum rule 
$\int_{0}^{\infty} \frac{2}{\pi} \,{d\Omega}\sigma(\Omega)=
\frac{n e^2}{m}$, with $n$ the total (conduction and valence)
electron density and $m$ the bare mass. Extracting the portion pertaining 
to the $e_g$ electrons from the measured conductivity has been 
controversial. We will show below
how to do this using the information on spectral weights we derive 
below.
Here we concentrate on spectral weight defined in 
Eq.~(\ref{spec-wt}). The real part of the optical conductivity
(Eq.~\ref{opt-cond1}) can be written as  
\begin{eqnarray}
\sigma(\Omega) & = & e^2 \sum_{\sigma}
\int\, d\epsilon_{k} N(\epsilon_{k}) \varphi(\epsilon_{k})
\int\, \frac{d\omega}{\pi} 
\frac{f(\omega)-f(\Omega +\omega)}{\Omega} \nonumber \\
& \times & A_{\sigma}(\epsilon_{k},\omega)A_{\sigma}(\epsilon_{k},\Omega
+\omega) 
\label{opt-cond2}
\end{eqnarray}
where $A_{\sigma}(\epsilon_{k},\omega) = -\frac{1}{\pi}
Im G_{\sigma}(\epsilon_{k},\omega)$ is the spectral function, 
$N(\epsilon_{k})$ the density of states (Eq.~\ref{dos}) and
$\varphi(\epsilon_{k}) = (4t^2 - \epsilon_{k}^2)/3$ is the current 
vertex for the Bethe lattice. We obtain 
$\varphi(\epsilon_{k})$, the form of which for the Bethe lattice 
has not been given before (Ref.~\cite{Lange99} presents $\varphi$ for
the Gaussian density of states), by requiring that $K$ be given both by the
explicit integration of $\sigma$ (Eq.~\ref{opt-cond2}) and by the generally
valid expression \begin{equation}
K = \int\, d\epsilon_{k} \frac{d\omega}{\pi} f(\omega) \epsilon_{k}
N(\epsilon_{k}) Im G(\epsilon_{k},\omega)
\,.
\end{equation}
Requiring the two to be equal yields the differential equation 
\begin{equation}
-\frac{\partial}{\partial \epsilon_{k}} 
\left[N(\epsilon_{k}) \varphi(\epsilon_{k})\right] = 
N(\epsilon_{k}) \epsilon_{k}
\,,
\end{equation}
the solution of which yields $\varphi(\epsilon_{k})$.

\section{NUMERICAL METHODS}
We now outline the numerical methods used to solve Eqs.~(\ref{mfeq})
and compute physical quantities. For computational convenience, we 
rewrite the self-consistency Eqs.~(\ref{mfeq}) in
terms of $b_0$ and $b_1$ defined in Eq.~\ref{b0b1}. 
At $T>T_c$, there is no magnetic order and $b_1(\omega)=0$. At $T=0$ and
non-zero $J$, all spins are ferromagnetically aligned implying $P(\theta) 
\rightarrow  \delta(cos\theta -1)$, and one can analytically solve for the
$b_{0,1}(i\omega_{n})$ from Eq.~(\ref{mfeq}). We initialize
$b_{0,1}$  with the $T=0$ expressions, and solve Eq.~(\ref{mfeq}) on the
Matsubara axis by direct iteration. 
\begin{figure}[htbp]
\vspace*{-0.4cm}
\epsfxsize=3.3in
\epsfysize=2.75in 
\epsffile{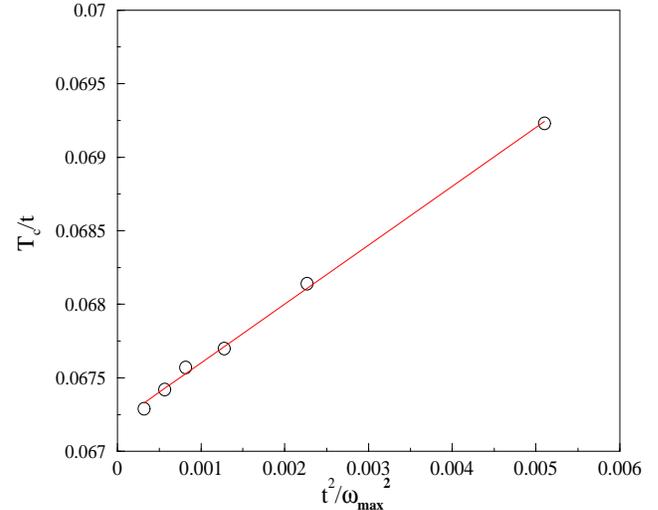}
\vspace*{-0.1cm}
\caption[]{{ The dependence of $T_c$ on the Matsubara cutoff 
$\omega_{max}$, for $n=0.75, n_{orb}=1$ and $J=10t$; $\omega_{max}$ ranges 
from $(4t+J)$ to $4(4t+J)$. The quadratic dependence on $t/\omega_{max}$ holds 
for the $n_{orb}=2$ as well. For our calculations we choose 
$\omega_{max}= 3(4t+J)$ which produces a  value of $T_c$ that is within
$0.5\%$ of the $\omega_{max} \rightarrow \infty$ value.}} 
\label{scaling}  
\end{figure}
It is convenient to compute $P(\theta)$ by 
solving Eq.~(\ref{mfeq}) on the Matsubara points, and then use this converged
$P(\theta)$ to solve the equations on the real axis. 
Convergence to within an rms error of $10^{-4}$ was
usually achieved within 12 iterations unless one is close to the magnetic
$T_c$ where there is critical slowing down. 
The electron density $n$ and the normalized magnetization density $m$ 
are given by  
\begin{eqnarray}
n & = & n_{orb} T \sum_{n} \left[G_{loc}^{\uparrow}(i\omega_{n}) + 
G_{loc}^{\downarrow}(i\omega_{n})\right] \\
m & = & \int_{-1}^{1} \,d(cos\theta) cos\theta P(\theta)
\label{mag}
\,.
\end{eqnarray} 
An accurate value of the transition temperature $T_c$ was most 
conveniently obtained by computing values of $m$ (Eq.~\ref{mag}) in the range 
$0.17\le m \le 0.3$ and finding $T_c$ by fitting to the mean-field 
expression $m^2 (T) = \alpha(T_{c}-T)$, with $\alpha$ and $T_c$ fit 
parameters.

The next issue is that of the number of Matsubara points needed to 
compute $S_{loc}(\theta)$, $n$ and $K$. 
At large $\omega_{n}$ , the asymptotics of the mean fields are
\begin{eqnarray}
b_{0} & = & i\omega_{n} + \mu - \frac{1}{i\omega_{n} + \mu} - 
\frac{1+J^2}{(i\omega_{n} + \mu)^3} \nonumber \\
b_{1} & = & \frac{Jm}{(i\omega_{n} + \mu)^2}  
\,.
\label{asymp}
\end{eqnarray}

In our computations we choose a frequency cutoff $\omega_{max}$ and
evaluate higher frequency contributions analytically using the 
asymptotic form given in Eq.~(\ref{asymp}). The errors in physical 
quantities are of order $\omega_{max}^{-2}$.
This is illustrated for $T_c$ in Fig.~\ref{scaling}. To achieve an 
accuracy of $5 \times 10^{-4}$ in $n$, it was required to choose 
$\omega_{max} \ge 3(4t+J)$. Choice of this value of
$\omega_{max}$ also ensures that the value of $T_c$ is within $0.5\%$ of the
$\omega_{max} \rightarrow \infty$ value.

\section{FERROMAGNETIC TRANSITION TEMPERATURE}

\subsection{Analytical results}

We first present mean field results and establish the phase diagram,
then show how to incorporate fluctuation corrections.
We begin with analytical results for small ferromagnetic
$J$. At $T=0$ the core spins are magnetically ordered with a
characteristic wavevector ${\vec q}$. The polarized
core spins produce an effective magnetic field on the conduction 
electrons which polarizes them, leading to a magnetization ${\vec m}_q$
and to a change in energy, which for small $J$ is
\begin{equation}
\delta E = \frac{1}{2} \left(\chi^{0}_{q}\right)^{-1} {\vec m}_{q}\cdot {\vec
m}_{-q}  + \frac{1}{2} J \left({\vec S}_{q}\cdot {\vec m}_{-q} + H.c\right)
\label{gstate}
\end{equation}
where $\chi^{0}_{q}$ is the magnetic susceptibility for noninteracting 
electrons at wavevector ${\vec q}$. Minimization of Eq.~(\ref{gstate}) leads
to ${\vec m}_{q} =  \chi_{q}{\vec S}_{q}$ and 
$\delta E = -\frac{1}{2} J^{2} S_{q}^{2} \chi_{q}$. Thus at small
$J$ the system will order at the wavevector which maximizes $\chi_{q}$.

\begin{figure}[htbp]
\vspace*{-0.6cm}
\epsfxsize=3.3in
\epsfysize=3.0in
\epsffile{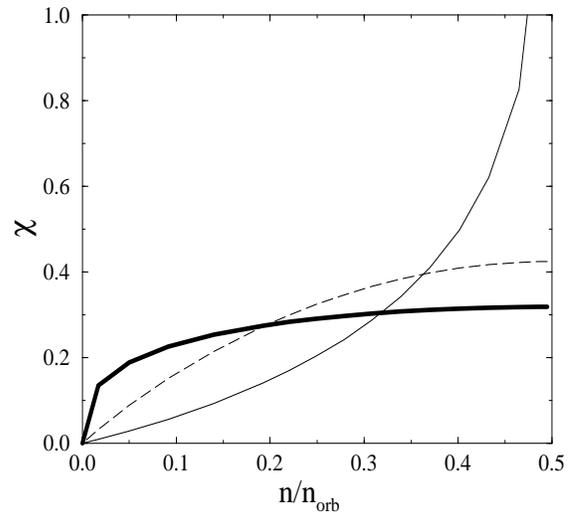}
\caption[]{{ The susceptibilites $\chi(q=0)$ (bold line), $\chi_{local}$
(dashed line) and $\chi(q=(\pi,\pi,....))$ as a function of the filling. 
The $J=0$ ground state is ferromagnetic where $\chi(q=0)$ is largest,
incommensurate where $\chi_{local}$ is largest, and commensurate
antiferromagnetic where $\chi_Q$ is largest. }}  
\label{chi}  
\end{figure}

The wavevector at which $\chi$ is maximal depends
on band filling and dimensionality. Here, 
for consistency and to facilitate comparison to other work
\cite{Furukawa,Yunoki98} we restrict ourselves to infinite-$d$. For
very large $d$, $\chi(q)$ is independent of $q$ except for regions of
width $O(1/\sqrt{d})$ about $\vec{q}=0$ and the commensurate antiferromagnetic
vector $\vec{q}=\vec{Q}=(\pi ,\pi , ...)$; therefore at $d=\infty$ we 
need to consider $\chi_{q=0}, \chi_{q=Q}$ and the susceptibility
at a typical $q$, $\chi_{loc}$\cite{georges96}.
Fig.~\ref{chi} shows the ferromagnetic (bold line),
antiferromagnetic (solid  line) and local (dashed line) susceptibilities for
$J=0$ calculated for a semicircular density of states. The phase 
corresponding to the maximal $\chi$ is the small $J$ ground state.
For $0\le n/n_{orb} \le 0.195$, 
$\chi(q=0)\ge\chi_{loc},\chi_Q$; this is the range of dopings where the 
DE model has a low $J$ ferromagnetic ground state.
At small $J$ and intermediate
$n$ ($0.195 <n/n_{orb}< 0.35$), $\chi_{loc}$ is largest, implying order
at wavevector different from both $0$ and $Q$. At $d=\infty$ all such 
wavevectors are degenerate, implying some sort of highly degenerate 
ground state. Finite $d$ corrections will select a particular wavevector. We
therefore identify the phase as incommensurate (IC). For $n/n_{orb} > 0.35$,
$\chi_Q$ is largest and the small $J$ phase is a commensurate antiferromagnet. 

Using mean field theory, we may also obtain an expression for $T_c$. 
For finite range interactions, 
mean field theory is strictly correct only in the limit $d\rightarrow 
\infty$. We will consider finite $d$ corrections below.
Writing Eq.~(\ref{gstate}) in real
space, focussing on a single  site $i$, integrating out the conduction
electrons and making a mean  field ansatz yields a self-consistent equation
for the polarization on site $i$. For example, for a ferro- or commensurate
antiferro-magnetic ordering, the effective field at site $i$, 
$h_{eff,i} = J \sum_{i\ne j} \chi_{i-j}\left<S_{j}\right>$, 
has magnitude independent of $i$ and we find
\begin{equation} 
T_{c} = \frac{J^2}{3} [\chi_{Q} - \chi_{loc}]
\label{Tc-ana1}
\,.
\end{equation}
An explicit analytic expression for
ferromagnetic $T_{c}$ can be obtained by working out the susceptibilities in
the mean-field  Eq.~\ref{Tc-ana1} for $Q=0$, and one finds that  
\begin{equation}
T_{c} = \frac{J^2}{9 \pi t} (\mu^2 -1) \sqrt{4-\mu^2}
\,.
\label{Tc-ana2}
\end{equation}
We observe $T_{c}\rightarrow 0$ at $\mu = 1$ which corresponds to 
$n=n_{c}=0.195$; for $n>n_c$, the small $J$ ground state becomes IC.

Next we consider the case of very large $J$, focussing first on 
the ground state energy. For $J>J^{*}(n)$, a ferromagnetic state 
would be fully spin polarized, and the ground state energy is simply 
that of the appropriate density of spin polarized electrons moving 
in the relevant band structure, and is therefore of order $t$. If
$n/n_{orb}=1$ the band is completely full, and the ground state 
energy vanishes, because no hopping is possible. For the semicircular 
density of states, $J^{*}(n_{orb})= 2t$ and if $n=n_{orb}(1-x)$ the
the ground state energy $E_{FM}$ is
\begin{equation}
E_{FM} = - 2 x t + O\left(x^{2} t\right)
\label{efm}
\,.
\end{equation}
At $n=n_{orb}$, a commensurate antiferromagnetic state is favoured, 
because virtual hops between the up and down sublattices leading to an energy
gain $\sim -t^{2}/J$ are possible. For the Bethe lattice with $Z$ nearest
neighbours and near neighbour hopping $t_\delta$, the energy is
\begin{equation}
E_{AF} = - \frac{2 Z t_{\delta}^2}{J}\rightarrow
- \frac{2 t^2}{J} - O\left(x^{2} t\right)
\label{eaf}
\end{equation}
where the arrow indicates the $d\rightarrow \infty$ limit. Further, one
may consider an incommensurate state which for the purpose of 
$d =\infty$ energetics has a random spin arrangement, leading to 
\begin{equation}
E_{IC} = - \frac{t^2}{J} - \sqrt{2} x t + O\left(x^{2} t\right)
\label{eic}
\,.
\end{equation}

\begin{figure}[htbp]
\vspace*{-1.2cm}
\epsfxsize=3.3in
\epsfysize=3.0in
\epsffile{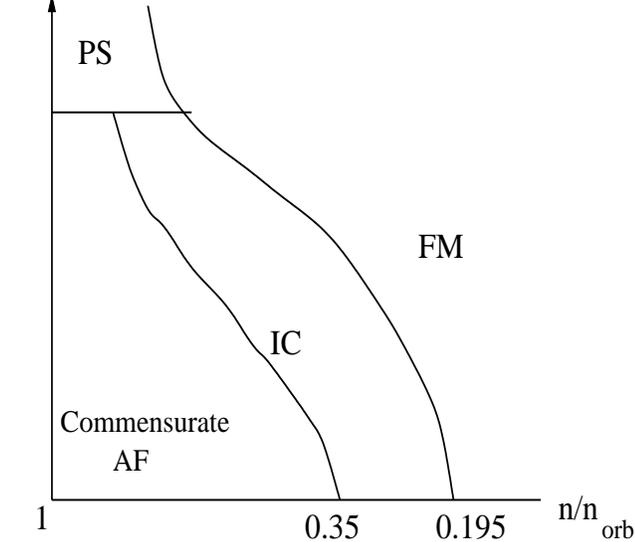}
\vspace{0.1in}
\caption[]{{ The phase diagram of the double exchange model, 
as deduced from the analytic arguments at small and large $J$. At small
$J$, one can have ferromagnetic (FM), commensurate anti-ferromagnetic
(AF) or incommensurate (IC) phase. At large $J$, the IC phase is not
energetically favoured and PS indicates phase separation between AF and FM. }} 
\label{phase}  
\end{figure}

Equating $E_{FM}$ to $E_{IC}$ implies a ferromagnet-incommensurate transition
at $x_{F-I} = t/\left((2-\sqrt{2})J\right)$ ; equating $E_{AF}$ to $E_{IC}$
yields incommensurate-antiferromagnetic transition at $x_{I-A} = t/(\sqrt{2}
J)$. Thus at small $1-n$ the sequence of spatially homogeneous phases as was
also found at small $J$ and small $n$. However, at small $1-n$ and large 
$J$, inhomogeneous phases are favoured: the standard Maxwell construction
applied to Eqs.~(\ref{efm},\ref{eaf},\ref{eic}) shows that in this model, which
neglects long ranged Coulomb interactions, the whole small $x$ and large $J$
regime is in fact phase separated into commensurate $AF$ and $F$ with $x\sim
1/2$,  as previously noted by \cite{Yunoki98}.
Combining the small $J$ and large $J$ results, we obtain the phase diagram
shown in  Fig.~(\ref{phase}). The qualitative behaviour is correct; we do not 
have precise numerical values for the IC-AF or PS phase boundaries.

We emphasize that
the large-$J$ antiferromagnetism and phase separated behaviour is tied to the 
regime $n\approx n_{orb}$, where the conduction band of the  ferromagnetic
state is almost completely full. The physically relevant  regime for the
manganites is  $n < n_{orb}/2$, where the large $J$  behaviour is simply
ferromagnetic. Several published papers\cite{Furukawa} have asserted that
behaviour associated with the regime $n\approx n_{orb}$ is relevant to the
manganites.  In our view, these
assertions are unjustified because $n_{orb}=2$, so the physical density 
corresponds to $n/n_{orb}<0.5$. 

It is possible that additional interactions, neglected here, could change the
effective orbital degeneracy from 2 to 1. We do not believe this happens in
the manganites.
In support of our view we cite the case of LaMnO$_3$. This is an insulator
which has a frozen Jahn-Teller distortion which acts to quench the 
orbital degrees of freedom. This material is a $(0,0,\pi)$ antiferromagnet:
its magnetic structure is ferromagnetic planes antiferromagnetically
coupled. The ferromagnetic in-plane coupling means that virtual hopping
to the empty orbital is the dominant process. The much weaker 
antiferromagnetic bond perpendicular to the plane is believed to be 
due to an additional $t_{2g}-t_{2g}$ superexchange which becomes 
important because the Jahn-Teller order strongly suppresses the out 
of plane hopping. In other words, even in the material in which the 
tendency to quench the orbital degree of freedom is strongest, there
is no conduction band mediated antiferromagnetism in the physically 
relevant regime. We believe, therefore, that the physics of $n\approx n_{orb}$
is simply not relevant to the manganites, in contrast to the assertions 
made in \cite{Yunoki98}.

In the same way, the phase separation discussed in
Refs.~\cite{Yunoki98} is crucially dependent on the existence of a
commensurate antiferromagnetic  order; the $(0,0,\pi)$ order observed in the
actual materials would not  lead to the same sort of phase separation, because
it would permit metallic in-plane conduction.

\subsection{Numerical results}

\begin{figure}[h]
\epsfxsize=3.3in
\epsfysize=3.0in
\epsffile{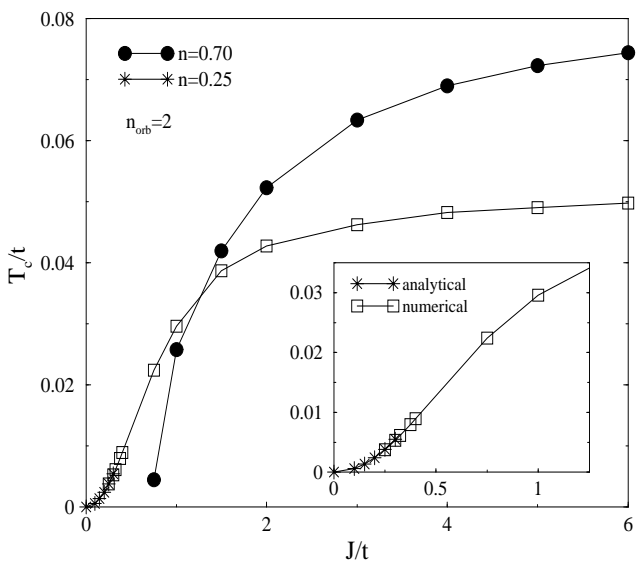}
\caption[]{{ $T_{c}$ vs $J$ plot of the double-exchange model for doubly 
 degenerate $e_g$ orbitals. The $n=0.7 ; n_{orb}=2$ curve is relevant 
 for La$_{0.7}$A$_{0.3}$MnO$_3$ compounds. For $n\le 0.39$, the model can
 sustain a ferromagnetic solution down to the lowest $J$.  The inset 
 amplifies the low $J$ end of the $n=0.25$ curve, to illustrate how
 the numerics (open boxes) patch on to the analytical expression (stars)
 from Eq.~(\ref{Tc-ana2}). The solid  lines are a guide to the eye.
 $T_c(J=\infty) = 0.079$ for $n=0.7$ and $0.055$ for $n=0.25$. }}   
\label{Tc-J}  
\end{figure}

By solving the dynamical mean field equations numerically, we have computed
the ferromagnetic transition temperature $T_c$ as a
function of the Hund's coupling. Fig.~\ref{Tc-J} shows ferromagnetic $T_c$
vs $J$ for fillings of $n=0.7$ (solid circles) and $n=0.25$ (squares), for
doubly degenerate $e_g$ orbitals ($n_{orb}=2; n/n_{orb}= 0.35$ or $0.125$). As
we noted in the analytic treatment of the low $J$ limit, for not too low 
densities ($n>0.2 n_{orb}$), a ferromagnetic solution cannot be sustained. Thus
for a modest $n$, as $J$ is decreased the ferromagnetic $T_c$
vanishes and the ground state changes from ferromagnetic to incommensurate
to antiferromagnetic. As noted previously, we find $n_{c}\approx0.195n_{orb}
\approx0.2n_{orb}$. We therefore suspect that the 
$n=0.2(n_{orb}=1)$ curve in the $T_{c}-J$ diagram of Ref.~\cite{Furukawa},
which indicates that $T_c$ drops to 0 at a finite $J$ of order 1, is
incorrect at the low $J$ end; the FM solution should be sustainable down to 
a very small $J$.
The stars in Fig.~\ref{Tc-J} correspond to Eq.~(\ref{Tc-ana2}), with the
chemical potential $\mu = -1.52t$ for $n=0.25; n/n_{orb}=0.125$(the curve is
magnified in the inset). We notice that one cannot sustain a ferromagnetic
transition for $n=0.7$ as $J\rightarrow 0$.

\begin{figure}[h]
\vspace*{-0.4cm}
\epsfxsize=3.3in
\epsfysize=3.0in
\epsffile{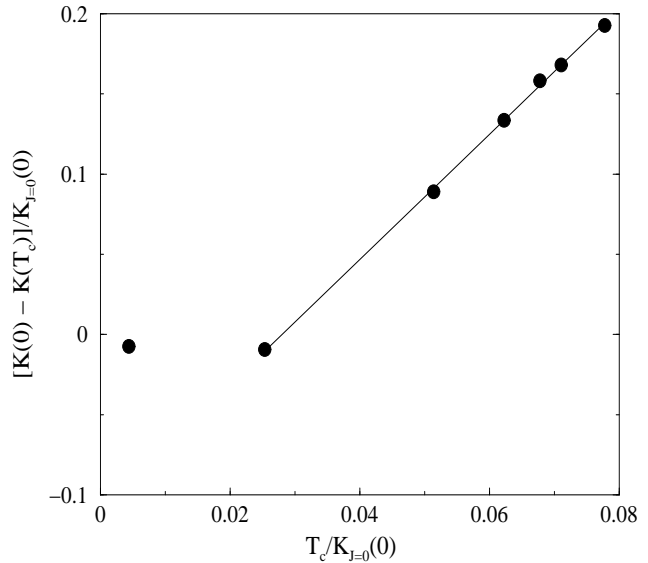}
\caption[]{{ $\Delta K = K(0)-K(T_{c})$ vs. $T_c$ for the DE
 model for $n=0.7; n_{orb}= 2$. Both the quantities are scaled with the $K$ for
the non-interacting system at $T=0$ ($K_{J=0}(T=0)= 1.01t$ for $n=0.7$).}} 
\label{KTc}  
\end{figure}

The ferromagnetic transition temperature has an apparently complicated 
dependence on interaction strength and doping. We now show that this 
seemingly complicated dependence simply reflects the variation of the 
kinetic energy, $K$, with these parameters. Fig.~\ref{KTc} plots $T_c$
against the change in kinetic energy between the paramagnetic and the 
$T=0$ ferromagnetic state ($\Delta K = K(0)-K(T_{c})$), both being 
normalized by the $K$ of the  noninteracting ($J=0$) system at $T=0$.

To understand this curve, consider the general expression for the free
energy, $F= E - TS$ of $H_{DE}$ (Eq.~\ref{Hdex}), where $S$ is the 
entropy and $E=-K-{\cal{J}}$ where $K$ is given by Eq.~(\ref{kin1}) and 
\begin{eqnarray}
{\cal{J }} & = & \sum_{i,\alpha,\beta} J \left<{\vec S}_{i} \cdot
d_{i\alpha}^{\dagger} \vec{\sigma}_{\alpha\beta} d_{i\beta}\right>\\ \nonumber
& = & T \sum_{n} ln\left[b_{n,\sigma} - J \sigma\right]
\,.
\end{eqnarray}
where the latter expression is the dynamical mean-field result.
Entropy favours the disordered state; ferromagnetic order is driven 
by a decrease in energy. $T_{c}>0$ implies $E_{fm}<E_{nm}$, where $E_{fm}$ 
is the ground state energy of the ferromagnet and $E_{nm}$ that of 
"non-ferromagnet"(incommensurate or antiferromagetic order).
At low $T$ all spins are aligned, and carriers can move freely, 
while at high $T$ the random spin arrangement and large $J$ means 
carrier hopping is somewhat blocked. $T_c$ is therefore set by the 
change in $K(T_{c})$, which is a simple number ($\approx 1/3$) of $K(T=0)$
as $J\rightarrow \infty$.

As $J\rightarrow \infty$, the local spin is always 
parallel to the core spin and ${\cal{J}}$ does not change between 
ferromagnetic and paramagnetic states. The transition is entirely 
driven by the change in $K$ between the paramagnetic and ferromagnetic 
states. In the non-ferromagnetic (NM) phase the band is narrower than
in the FM phase, but at finite $J$ the electron has some
possibility to hop onto the "wrong" spin site. This is equivalent to the
on-site magnetization not  being saturated in the NM phase, thus
${\cal{J}}_{fm} > {\cal{J}}_{nm}$. Since $T_{c}\sim E_{nm}-E_{fm} = \Delta K +
\Delta {\cal J}$, it implies that $E_{fm}=E_{nm}$  when $T_{c}=0$ and we get
$K_{fm}<K_{nm}$ at this point. So at high $J$, we start with fully polarized
bands and with the magnetic state having more kinetic energy. The high $J$
expansion suggests that everything comes from virtual hops : $\Delta K \sim
t^{2}/J$ and $\Delta {\cal J} \sim J\Delta n \sim J\cdot t/J \sim t$. Since
$\Delta {\cal J}$ is independent of $J$ at large $J$,$T_c$ is linear in
$\Delta K$ for large values of $J$. As we reduce $J$ to a value where the
polarization isn't complete, first $K$'s cross and then $T_c \rightarrow 0$.
We have used the dynamical mean-field equations to determine the points 
at which $E_{fm}=E_{nm}$ in Fig.~\ref{phase}. For $n = 0.7 (n/n_{orb}=0.35)$
the critical $J \approx 0.6t$.

Since $K$ is of crucial interest to us, we present its values at
$T=0$ for the soluble limits $J=0,\infty$ in Table I for a variety of
fillings. At $J=0$ the system is always paramagnetic, and the kinetic 
energy increases with $n$ in the range $0<n<1$. For $J=\infty$ 
however, the kinetic energy is maximal for $n=n_{orb}/2$. As 
$K(T_{c})= K(0)/\sqrt{2}$ for $J=\infty$ and the transition temperature 
is tied to the kinetic energy (Fig.~\ref{KTc}), $T_c$ also is a maximum at
the same filling, as has been noted in earlier works\cite{Millis96b,Furukawa}.
\begin{minipage}{3.4in}
\begin{table}
\caption{The kinetic energy $K$ at $T=0$, evaluated using a dynamical mean
field method with a semicircular density of states, for $J=0$ and $J=\infty$ at
different fillings.}  \label{table1}
\begin{tabular}{cccl}
$n/n_{orb}$&$K(J=0)/n_{orb}$&$K(J=\infty)/n_{orb}$\\
\hline\hline
0.125&0.2160&0.1958\\ \hline
0.25&0.3916&0.3248\\ \hline
0.35&0.5086&0.3889\\ \hline
0.50&0.6496&0.4244\\ \hline
0.75&0.7994&0.3248
\end{tabular}
\end{table}
\end{minipage}  

We have argued that the kinetic energy is the crucial parameter determining
$T_c$. To further substantiate this we show in Table II results obtained 
for $T_c(J\rightarrow \infty)$ by 
a variety of techniques in a range of models\cite{tokurabook}, expressed in 
terms of $K_{J=\infty}(T=0)$. Roder et al.\cite{Roder97} used a series 
expansion technique.
Yunoki et al.\cite{Yunoki98} have
studied thermodynamic properties of the classical core-spin model in
$d=3$ with a single orbital nearest neighbour hopping using Monte
Carlo on 6$\times$6$\times$6 clusters. Calderon and Brey\cite{Calderon98} 
have performed similar Monte Carlo calculations on 
4$\times$4$\times$4 and 20$\times$20$\times$20 lattices, and
argued that Yunoki et al.\cite{Yunoki98} underestimated $T_c$ by a
factor of $\sim 1.6$. The results of Calderon et al. are in agreement with
those of Roder et al. Note that in all calculations the doping dependence 
of $T_c$ is essentially the doping dependence of the kinetic energy. 

We now consider the dynamical mean field results. These yield somewhat 
higher $T_c/K$ results, as expected because fluctuations are neglected. 
Calderon and Brey partitioned the hopping term $t_{ij}$ into average 
($\bar{t}$) and random ($\delta t_{i,j}$) components and argued that 
fluctuations lead to a $\approx 25\%$
correction to $T_c$. Applying this correction leads to the numbers given in 
brackets, which are in good agreement with those found in
Refs.~\cite{Calderon98,Roder97}. 
\begin{minipage}{3.4in}
\begin{table} 
\caption{ $T_{c}/K(T=0)$ for $J=\infty$; comparison of the different 
methods.The bracketed terms for $d=\infty$ indicate 
$T_{c}$ values that are $\sim 25\%$ fluctuation corrected. }
\label{table2}
\begin{tabular}{cccl}
$Method$&$n/n_{orb}$&$T_{c}/K_{J=\infty}(T=0)$\\
\hline\hline
d=$\infty$&0.5&0.070(0.053)\\ \hline
d=$\infty$&0.25&0.063(0.050)\\ \hline
Series\cite{Roder97}&0.5&0.053\\ \hline
Series\cite{Roder97}&0.25&0.050\\ \hline
Yunoki\cite{Yunoki98}&0.5&0.036\\ \hline
Yunoki\cite{Yunoki98}&0.25&0.033\\ \hline
Calderon\cite{Calderon98}&0.5&0.056\\ \hline
Calderon\cite{Calderon98}&0.25&0.056
\end{tabular}
\end{table}
\end{minipage}

\section{OPTICAL CONDUCTIVITY}
\begin{figure}[htbp]
\vspace*{-0.5cm} 
\epsfxsize=3.3in
\epsfysize=1.5in 
\epsffile{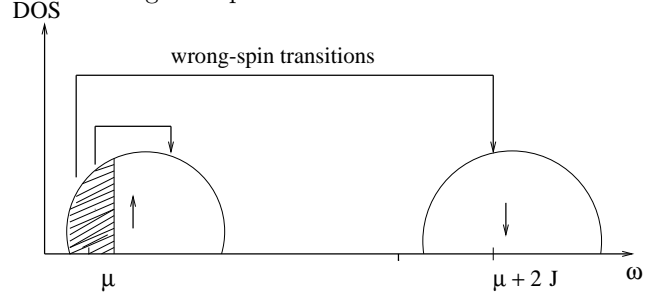}
\vspace{0.1in}
\caption[]{{ The large $J$ density of states, consisting of bands of 
electrons parallel ($\uparrow$) and antiparallel ($\downarrow$) to the
core spin along with possible optical transitions indicated by arrows. 
$\mu$ denotes the chemical potential, and the  shaded area represents the
electron filling. There are two pieces to  $\sigma(\omega)$; the transitions
within the lower band with  $\omega \sim t$ and the "wrong-spin" transitions
from the lower to the upper spin band around $\omega \sim 2J$. The
"wrong-spin" transitions become allowed as $T$ is increased from $0$. For
large $J$, these two pieces are well separated. }}  
\label{wrong-spin}   
\end{figure}

Next we discuss the optical conductivity of the double-exchange model,
At large $J$ the density of states of this model consists of two nearly
semi-circular bands corresponding to conduction electrons parallel
($\uparrow$) and antiparallel ($\downarrow$) to the core spin. The bands
are separated by an energy $2J$. This structure is shown in
Fig.~(\ref{wrong-spin}), which we now use to give a qualitative discussion of
$\sigma$. 
The crucial point is that the optical process conserves electron
spin. At $T=0$ in the fully polarized
ferromagnetic state all spins are aligned, making it impossible for an
optical process to create a final state with an  $e_g$ electron anti-aligned
to a core spin. Further, the perfect spin alignment means that no scattering
processes are present; thus the $T=0$ conductivity is simply given by the band
theory of the spin polarized $e_g$ electron manifold.
The spectral weight in these transitions  follows directly from the band theory
kinetic energy. Impurities, weakly coupled phonons etc. will change the form
of the conductivity but will  not significantly affect the total spectral
weight. 

As $T$ is raised, the spin  disorder increases, and an electron moved to an
adjacent site may find itself anti-aligned to the new core spin
\cite{Okimoto95}.

\begin{figure}[htbp]
\vspace*{-1.3cm}
\hspace*{-0.2cm}
\epsfxsize=3.3in
\epsfysize=3.3in 
\epsffile{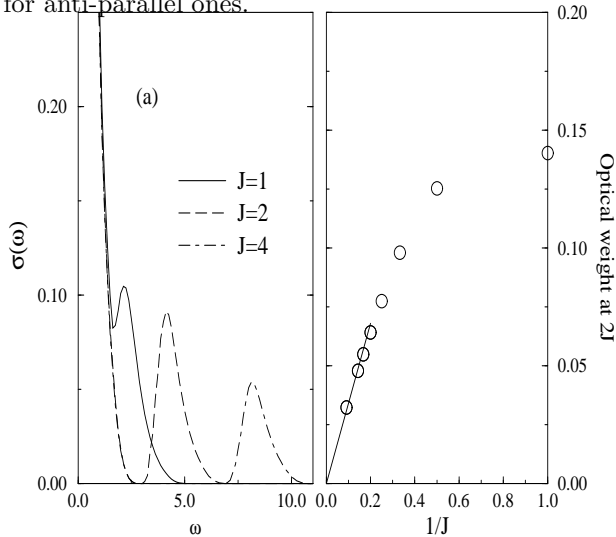}
\vspace{0.1in}
\caption[]{{ (a) The optical conductivity at $T=T_c$ for the doubly
-degenerate DE model for $n=0.7$. The results shown are for $J$ =1,2 and 
4; the inter-band peak is at $2J$. (b) Spectral weight in the
$e_{g}-e_{g}$ inter-band transitions, for $n=0.7$ and $T=T_c$. For
$J>>t$, the inter-band weight falls off as $\frac{n}{J}$. The straight line
has a slope of $n/2$, corresponding to the large $J$ analytic 
expression Eq.~(\ref{int-slope}). }} 
\label{sigma1}  
\end{figure}

Thus as $T$ is raised the resulting spin disorder leads to a broadening of
$\sigma$, a decrease in total spectral weight, and also to a shift of
oscillator strength to the peak at $\sim 2J$. For $2J>>t$, the wrong-spin
peak is well separated from the parallel spin transitions and we shall derive
its spectral weight in this limit. Fig.~\ref{sigma1}(a) shows the real part of
the optical conductivity (Eq.~\ref{opt-cond2}), calculated within the
dynamical mean field theory, for the experimentally relevant filling of
$n=0.7$, for doubly degenerate conduction electrons. The results shown are for
$J/t=1,2$ and $4$. The wrong-spin transitions produce the peak at $2J$, for
$J>1$ this becomes  well separated from the same-spin transitions.

We are interested here in what fraction of the total optical weight goes into
the wrong-spin transition. The spectral weight ratio is insensitive to details
of the band structure.
In what follows, we give an analytical estimate for the oscillator strength at
$2J$ for large values of $J$ and compare it with our numerical results. 
For $J>>t, \Omega \sim 2J$ and $T=T_c$, the inter-band optical weight is 
\begin{eqnarray}
\int\, \frac{2}{\pi} d\Omega \sigma(\Omega) & = &
\int\, d\epsilon_{k} N(\epsilon_{k})\varphi(\epsilon_{k})
\int^{\mu}_{-\infty}\, \frac{d\omega}{\pi} A_{-}(\epsilon_{k},\omega)
\nonumber \\
& \times & \int^{\infty}_{-\infty}\, \frac{d\omega^{\prime}}{\pi} 
\frac{A_{+}(\epsilon_{k},\omega^{\prime})}{J}
\end{eqnarray}
where we have used Eq.~(\ref{opt-cond2}) for $\sigma$. Here, $A_{-}$ is the
imaginary  part of the Green function for parallel electrons and $A_{+}$ for
anti-parallel ones. 

Now, $A_{+}$ is an analytic function and its integral gives the 
real part evaluated at $\omega=J$, which is $1/(2J)$ in the 
large $J$ limit. Thus
\begin{equation}
\int\, \frac{2}{\pi} d\Omega \sigma(\Omega) = 
\int\, d\epsilon_{k} \frac{(4-\epsilon_{k}^2)^{3/2}}{12\pi J} 
\int^{\mu}_{-\infty}\, \frac{d\omega}{\pi} A_{-}(\epsilon_{k},\omega)
\,.
\end{equation}
The $\omega$-integral gives the momentum space occupancy 
$n(\epsilon_{k})$. If the scattering is strong enough, this is just $n$.
So, for large $J$ and at the magnetic transition we get
\begin{equation}
\int\, \frac{2}{\pi} d\Omega \sigma_{2J}(\Omega) = \frac{n}{2J}
\label{int-slope}
\end{equation}
which falls off as $1/J$ and is proportional to the filling $n$. The 
numerical results are shown in Fig.~\ref{sigma1}(b); we see that 
the analytic value for the slope, $n/2=0.35$ (straight line) from
Eq.~\ref{int-slope} overlays the numerical curve for large values of $J$.
This analysis shows that the spectral weight in the wrong-spin 
transitions depends crucially on $n$. A previous calculation of $\sigma$
used an $n_{orb}=1$ and $n=0.7$. The resulting oscillator
strength in the wrong-spin transition is therefore not applicable to
manganites with $n/n_{orb}= 0.35$ (and indeed the oscillator strength 
in \cite{Furukawa} is larger than the one we calculated for the physical
$n/n_{orb}$ by a factor of two). To aid in comparison to experiments we give in
Table.~\ref{table2b} the total $T=0$ spectral weight, the total  $T>T_c$
spectral weight and, for $T>T_c$, the spectral weight in the  same-spin and
wrong-spin transitions, calculated using the dynamical  mean field theory for
various values of $J$.  
\begin{minipage}{3.4in}
\begin{table} 
\caption{ The total $T=0$ spectral weight $K(0)$, the total 
$T>T_c$ spectral weight $K(T_c)$ and , for $T>T_c$, the spectral weight in
the wrong-spin ($K_{anti}$)and same-spin ($K_{par}$) transitions for various
values of $J$, and $n=0.35 n_{orb}$.} 
\label{table2b}
\begin{tabular}{cccccl}
$J$&$K(0)$&$K(T_c)$&$K_{anti}$&$K_{par}$\\ \hline\hline
1&0.7778&0.8060&0.1209&0.6850\\ \hline
2&0.7778&0.6994&0.1067&0.5926\\ \hline
3&0.7778&0.6541&0.0833&0.5707\\ \hline
4&0.7778&0.6291&0.0661&0.5629\\ \hline
5&0.7778&0.6136&0.0547&0.5588\\ \hline
6&0.7778&0.6030&0.0467&0.5562\\ \hline
7&0.7778&0.5952&0.0408&0.5543\\ \hline
11&0.7778&0.5778&0.0275&0.5502\\ 
\end{tabular}
\end{table}
\end{minipage}

\section{COMPARISON WITH EXPERIMENTS}

\subsection{Overview}
In this section we relate our results to experiments. The measurements we
analyse are the value of the ferromagnetic transition temperature and the
magnitude and temperature dependence of the spectral weight (integrated 
area) in different regions of the optical conductivity spectrum. The 
physics issues we are able to clarify include the extent to which the
double-exchange-only model describes the behaviour of the manganites, 
the proper interpretation of the optical spectrum, and the value of the
Hund's coupling. The remainder of this section is organized as follows.
In sub-section B we summarize our results in a manner suited to comparison
to data, in sub-section C we outline what is known about the optical 
conductivity data, emphasizing the heretofore unresolved issues arising in 
interpretation of experiments and in sub-section D we present the comparison
between our results and data.

\subsection{Summary of theoretical results}

We have studied the double-exchange-only model defined by Eq.~\ref{Hdex}. 
This model involves itinerant electrons hopping among sites of a lattice
and coupled ferromagnetically to electrically inert core spins. No other 
interactions are explicitly included. This model captures some aspects of
the CMR manganese perovskites and of other 'double-perovskite' systems
\cite{KobayashiNature}; whether other physics is important is the subject of 
present debate\cite{Millis96b,DagottoScience}.

The double-exchange-only model was shown to have two crucial parameters:
the electron kinetic energy $K$ defined in Eq.~\ref{kin1} and the 
itinerant electron-core spin coupling $J$. $K$ is the expectation value
of a local operator. It depends on parameters like temperature
and $J$, but is insensitive to details of band structure. For the $J$ values
and carrier concentrations of physical relevance, the ground state is a fully
polarized feromagnet and $K(T=0)$ is independent of $J$ and in the
double-exchange-only model may be computed via a simple band structure
calculation. The value appropriate to La$_{0.7}$Sr$_{0.3}$MnO$_3$ is
$K\approx 0.84 eV$. We calculate $T_c$ as a function of
$K(T=0)$ and $J$ using a mean field method which treats local dynamics exactly
\cite{georges96}, and by comparison to published Monte Carlo calculations
\cite{Calderon98} were able to estimate the corrections to $T_c$ from 
non-mean-field spatially dependent fluctuations. The dynamical mean field
results are often published in terms of a bandwidth paramete $t$; the $t$ 
corresponding to the band theory $K$ is $t\approx 1.07 eV$.

By comparing our results to those obtained by other techniques, we found 
that in the $J\rightarrow \infty$ limit $T_{c}/K(T=0)\approx 0.16$, 
essentially independent of model details. For finite $J$, we found (for
carrier concentrations relevant to the CMR materials with ferromagnetic
ground states) that $T_c$ was linearly proportional to the change, 
$\Delta K$, of $K$ between $T_c$ and $T=0$. We further found that as 
$J\rightarrow \infty$, $\Delta K/K\rightarrow 1/3$; as $J$ is decreased,
$\Delta K$ decreases. At a critical $J$, $\Delta K =0$ and below this $J$
a ferromagnetic ground state cannot be sustained.

As previously noted, $K$ is measurable in optical experiments, and therefore
a quantitative test of the double-exchange-only model is possible and 
conversely constraints on optical conductivity may be extracted from 
measured $T_c$'s. The issues involved in this analysis will be treated in 
the next sub-section. To conclude the present sub-section we discuss the
additional information concerning the strength of non-double-exchange
interactions  which may be obtained from the comparison of $T_c$ and the
measured $K(T)$.

We may represent the electronic energy of a solid, $E_{el} = -K + I$ 
where $K$ was defined in Eq.~\ref{kin1} and $I$ represents the expectation
value of all interactions, including the $J$ term and other interactions
not included in the double-exchange model. In the rest of the discussion 
we assume $J$ is large enough to have a spin-polarized ground state.

In the double-exchange-only model at $T=0$, $K$ is saturated and given 
by the band theory value (assuming spin-polarized electrons). As $T$ is
increased, the entropy driven spin-disorder leads to a reduction in $K$; 
the change in $K$ between $T=0$ and $T>T_c$ depends on $J$, and becomes 
as large as $1-1/\sqrt{2} \sim 30\%$ in the $J\rightarrow \infty$ limit.
Other interactions do not commute with $K$, and therefore change the 
electronic state in a way which reduces $K$. Thus, we argue that
$K$ cannot be greater than the spin-polarized band theory value, and that 
a $K(T=0)$ appreciably less than this value indicates other interactions 
are important.

As temperature is increased from $T=0$, spin disorder leads to a decrease
in $K$ and therefore to an increase in the relative strength of the
interaction terms which in turn causes a further decrease in $K$. This
self-consistent effect says that for a given $J$, the relative change
$\Delta K/K$ in kinetic energy between $T_c$ and $T=0$ increases with 
increasing interaction strength. This physics was investigated in 
Ref.~\cite{Millis96b} for $J=\infty$ in the particular case of electron-
phonon interaction, but is expected to be more general. Further, because
kinetic energy is what decides $T_c$, the interaction induced decrease in
$K$ must decrease $T_c$ below the value predicted by double-exchange.

To summarize, the double-exchange-only model predicts a definite set of
relationships between $T_c$, $K(T=0)$, and the $T$-dependence of $K$ and 
$\Delta K$. These are summarized in Fig.~\ref{KTc} and Tables.~\ref{table2b}
and \ref{table3}.

 Adding other interactions causes $T_c$ and $K(T=0)$ to decrease 
and $\Delta K$ to increase.

\subsection{Optical conductivity}

This subsection discusses the interpretation of optical conductivity
data. As we have shown, the magnitude and temperature dependence of the 
spectral weight in the $e_g$ contribution to the optical conductivity 
contains crucial information about the electron kinetic energy and 
Hund's coupling. In order to extract this information one must identify 
the $e_g$ contribution to the measured conductivity, which is not 
straightforward because of the overlap between the $e_g$ transitions of
interest and processes involving other bands. In this subsection we analyze
the issues arising in the interpretation of the measured conductivity, with
emphasis on how our theoretical results can be used to resolve some of the
difficulties in interpreting data. 

The optical conductivity $\sigma(\omega)$ is the linear response 
function relating spatially uniform frequency dependent current
$\vec{j}(\omega)$ to applied electric field $\vec{E}(\omega)$. In simple 
terms, the conductivity d escribes how electrons move in response to an
electric field, and thereore contains information about interactions
which may hinder this motion. In metals it is useful to distinguish
between intra-conduction-band processes (those which involve scattering
of electrons between conduction band states, which in many cases including
the CMR manganites are the states of immediate physical interest) and 
other processes which involve scattering of electrons from conduction 
band states to other (empty) bands, from other (filled) bands to the 
conduction bands, or which do not involve the conduction bands at all. The
other processes are usually called interband, but in the manganites there are 
two orbitals per unit cell and therefore some of the intra-conduction-band
processes are, strictly speaking, interband.

In any event, one would like to extract from the measured conductivity
the portions pertaining to transitions between the states of interest (in
the manganite case, the intra-conduction-band processes) and analyse only 
these. However in many cases involving transition metal oxides it has not 
been clear how to separate the interesting conduction band contributions 
from other un-interesting processes. The CMR  materials are a promising 
system in which to investigate this issue because the conduction band 
contribution to $\sigma$ has a strong temperature dependence and a definite
relation to $T_c$, which can be used to distinguish it from other
contributions.

The main point is this: the double-exchange-only model predicts a 
definite relation between the $T=0$ $e_g$ oscillator strength and $T_c$.
Interaction corrections only reduce $T_c$ below the double-exchange value 
and $K$ below the band theory value. Therefore the part of the optical 
spectrum asigned to the $e_g$ electrons must contain at least enough 
kinetic energy to reproduce the observed $T_c$, but cannot contain more 
kinetic energy than predicted by spin-polarized band theory results.

A further constraint is provided by the change $\Delta K$ in kinetic 
energy between $T=0$ and $T_c$. For $H_{DE}$ (Eq.~\ref{Hdex}),
$\Delta K \le 0.3 K(T=0)$ and the maximal value occurs as $J\rightarrow
\infty$. Limited information is available concerning models with additional
interactions, but published calculations for the double-exchange plus
phonon problem at $J\rightarrow \infty$\cite{Millis96b} show that the
fractional change $\Delta K/K(T=0)$ can become slightly larger than 0.5, 
but the magnitude of $\Delta K$ is never much larger than the double-
exchange-only value. This is useful because the change in optical spectral
weight can be accurately measured, and the observed changes can with 
confidence be attributed to the $e_g$ electrons of interest. A final
constraint comes from the position and spectral weight of the "wrong-spin"
transitions. The point is that as
$T\rightarrow 0$, all of the spins are aligned in both the ground 
state and all states accessible from it via the optical matrix element. 
However, for $T>T_c$, the core spins are completely disordered and therefore
when an electron hops (or is pushed via the optical matrix element) from 
one site to another it has a probability 
for landing in the "wrong-spin" configuration,i.e. with $e_g$ electron 
anti-parallel to the core. The probability depends on the hopping matrix
element (i.e, the $T=0$ kinetic energy) and the value of $J$. This physics
was first pointed out by Okimoto et.al\cite{Okimoto95}, who, however, 
obtained what we argue below was an incorrect value of $J$.
We have determined the optical oscillator strength in this transition, 
as a function of $K(T=0)$ and $J$. The oscillator strength has a strong
dependence on carrier density not noticed in previous work.

\subsection{Analysis of data}

Optical conductivity has been measured in a wide range of manganites by
several groups\cite{Okimoto95,Okimoto97,Noh98,Noh99}. Results are
qualitatively similar, but there are substantial quantitative differences
between results of different groups. Some of the differences seem to be
experimental artifacts associated with surface
preparation\cite{Takenaka,Noh99}; others seem to relate to
differences in physics among various members of the CMR family of 
materials. As we shall see, our results provide consistency checks 
which allow one to separate artifacts from intrinsic behaviour and to make
some statements about the underlying physics.

To be concrete, we discuss the data of Quijada et.al\cite{Quijada98}, 
who measured $\sigma(\omega,T)$ for three pseudo-cubic manganite films
La$_{0.7}$Sr$_{0.3}$MnO$_3$ (LSMO), 
La$_{0.7}$Ca$_{0.3}$MnO$_3$ (LCMO) and Nd$_{0.7}$Sr$_{0.3}$MnO$_3$ (NSMO).
The qualitative features of the data are (i) at $\omega \le 3eV$, an
absorption with a pronounced frequency dependence and an intensity that
shifts to lower frequency and increases markedly as $T$ is decreased
from $T_c$ to a low temperature, (ii) a strong feature centered at 
$\omega \approx 4eV$ with little apparent $T$-dependence, and (iii) a weak 
feature at $\omega \approx 3eV$, visible as a decrease in absorption 
intensity as $T$ is decreased below $T_c$. The interpretations offered 
in Ref.~\cite{Quijada98} was that (1) the integral of the low $T$ 
conductivity between $\omega =0$ and $\omega =2.7 $eV was a good 
representation of the total low $T$ conduction-band spectral weight, 
(2) that the strong feature at $\omega \approx 4eV$ was a Mn-O interband
transition and (3) the weak feature at $\omega \approx 3eV$ was the "peak 
at $J$" (it is actually at $2J$ in our units, as in Fig.~\ref{sigma1}) 
due to "wrong-spin" transitions characteristic of the spin  disordered state.
This identification is controversial. Okimoto et al.\cite{Okimoto95,Okimoto97}
identified a lower energy feature as the peak at $J$.

We begin our discussion with the "peak at $J$". The identification
offered by Okimoto et al.\cite{Okimoto95,Okimoto97}, implies
$J\approx 0.75 eV$ which is less than the dynamical mean field
hopping $t=1.08 eV$ (this is derived from equating the dynamical mean field 
kinetic energy $K$ to the band theory $K = 0.84meV$). Thus the $J$ must 
be larger as a ratio of $J/t\approx 0.7$ would imply a ferromagnetic state 
cannot be sustained at $n/n_{orb}=0.35$, in contradiction to experiment. 

We now turn to the identification of Quijada et al.\cite{Quijada98}, which
corresponds to $J=1.5eV$, i.e to $J/t \approx 1.38$ or a fluctuation corrected
$T_c \approx 0.03t\approx 310 K$, somewhat below the 350K observed for
La$_{0.7}$Sr$_{0.3}$MnO$_3$. This casts doubt on the interpretation offered by
Quijada et al. Further, the spectral weight observed by Quijada et
al. in the "peak at $J$", $K_{anti} = 34$meV is
about twice the spectral weight predicted by our calculation, using the
Quijada $J$ and the $t$ inferred from band theory. We conclude that the
feature  observed by Quijada et al. is  unlikely to be the "peak at $J$" and
that the true Hund's coupling is probably larger so that the
peak at $J$ is outside the experimental range. At least one
needs a value of $J\approx 2eV$ to reproduce the observed $T_c$ using the band
theory $K$. 

We digress briefly on the question of the origin of the observed feature.
It seems reasonable that it is caused by a shift of the energy of the Mn-O
interband  transition at 4eV, due to changes in the $e_g$ band as $T$ is
varied  through $T_c$ (the leading edge of this transition is an excitation
from a filled oxygen state to the Fermi level in the $e-g$ band). One
possibility is the double-exchange induced shift in chemical
potential, discussed in a different context in \cite{Furukawa}. Unfortunately,
for relevant dopings, the changes that we find for $J/t >2$ are of the
order of $1-2\%$, too small to explain the observed feature. An alternative 
possibility is a polaronic shift asociated with a  change in effective
electron-phonon interaction\cite{Millis96b,Millis96a}, but this has
not been studied in detail.

We now present a more detailed discussion of the remainder of the
spectrum, proceeding on the assumption that the "peak at $J$" is 
not visible in the spectrum at $\omega < 4eV$, i.e $J\ge 2eV$ and that 
the total $T=0$ $e_g$ spectral weight is less than or equal to the
band theory value $K=280meV$. 
Referring to Fig.~6 of Quijada et al.\cite{Quijada98}, one observes a
change between $T_c$ and $T=0$ $\Delta K(\omega^{\ast}) =
\int^{\omega^{\ast}}_{0} \, d\omega \sigma(T\rightarrow 0,\omega) - \sigma(T>
T_{c},\omega)$ in spectral  weight of 0.1-0.13 eV in LSMO, 0.1-0.12 eV in LCMO
and 0.1 in NSMO.  Consider LSMO first: if a double-exchange only model
described the  physics, the total $T=0$ spectral weight would be at least
$3\Delta K$,  i.e $K(T=0) > 0.3 eV$. This is slightly greater than the band
theory value. Now the $T_c$ of LSMO is $\approx 350 K\approx 0.1 K(T=0)$. 
Reproducing this $T_c$ with $K(T=0)=0.3 eV$ would require a $J$ slightly
less than 2eV.  From Table.~\ref{table2b} we see that such a $J/K$ would
imply a change $\Delta K$ in spectral weight much smaller than observed.
Similar, but more severe problems arise for the other materials, LCMO and
NSMO. The double-exchange-only constraint $K(T=0)\ge \Delta K$ implies 
$K(T=0) \ge 0.3 eV$ for these materials as well. The lower $T_c$'s 
($\approx 270 K$ for LCMO, $\approx 250 K$ for NSMO) would then imply 
$J=1.5-1.8 eV$, again yielding a $\Delta K$ that is too small, and a 
"peak at $J$" visible in the spectrum.

We would like to add here that our values of $K$ are a factor of 3
higher than that of \cite{Quijada98}. This is because of the $1/\sqrt{d}$
dependence of each current vertex in the expression for $\sigma(\omega)$. 
Thus, for $d=3$, the kinetic energies of Quijada et al should be 
multiplied by $(\sqrt{d})^{2}=3$ to compare with our numbers. We do this 
in Table~\ref{table3} for facilitate comparison.

To conclude this section we invert the logic, using the observed 
transition temperatures and $\Delta K$ to obtain bounds and estimates for
the conduction band spectral weights. We argue that for a given $K(T=0)$
and $J$, the double-exchange-only model gives an upper bound for $T_c$. 
Thus values of $T_c$ and $J$ yield a lower bound for $K(T=0)$. These 
bounds are shown in the Table~\ref{table3} for $J=2eV$, $3eV$ and $4eV$ 
for the systems studied by Quijada et al. Note that an upper bound of
$K(T=0)$ is given by the band theory value.

An alternative estimate may be obtained by considering the limit 
$J \rightarrow \infty$. In this case, as noted in \cite{Millis96b}, 
$T_c$ is a universal function of $K(T_c)$ (which may itself be affected 
by other interactions). Further, $T_c$ must decrease as $J$ is
decreased; thus, we may obtain a lower bound on $K(T_c)$ from the 
$J=\infty$ result for $T_c$. These bounds are also listed in 
Table~\ref{table3}.
\begin{minipage}{3.4in}
\begin{table} 
\caption{ The lower bounds on the kinetic energies ($K_{min}$) at $T=0$ and
$T_c$ for the systems studied by Quijada et al., shown for $J= 2eV, 3eV$ and
$4eV$. The upper bound is given by the band theory value $K_band(0)=840 meV$.
To compare with experiments, we cite the values of Ref.~\cite{Quijada98} in
the last two rows. The experimental values for $K$ are represented in the
conventions of this paper, i.e. values from \cite{Quijada98} multiplied by 3.
All energies are in meV. } 
\label{table3}  
\begin{tabular}{cccl}
$$&LSMO&LCMO&NSMO\\ \hline\hline
$K_{min}(0); J=2$eV&630&392&350\\  \hline
$K_{min}(0); J=3$eV&535&336&327\\  \hline
$K_{min}(0); J=4$eV&514&345&314\\  \hline
$K_{min}(T_c); J=\infty$&477&330&309\\ \hline
$K_{expt}(0)$&780&660&600\\ \hline
$K_{expt}(T_c)$&477&393&390\\
\end{tabular}    
\end{table}
\end{minipage}

It is interesting to note that the experimental $K(T_c)$ for LSMO is 
477meV, which exactly saturates the lower bound on $K(T_c)$ from double-
exchange. Because the infinite-$J$ double-exchange-only $T_c$ is expected 
to be the upper
bound to the true $T_c$ of an interacting model we expect that
in the actual material the $J$ is larger than $2 eV$ (so the model is not
far from the $J = \infty $ limit and we suspect that the integration
up to $\omega=2.7eV$ does not capture quite all of the spectral weight.

\section{CONCLUSIONS}

We have given a complete and correct treatment of the phase diagram, 
kinetic energy and some aspects of the optical conductivity of the 
double-exchange-only model of mobile carriers coupled to core spins. 
We have determined the physics operating in different regions of the
phase diagram and have demonstrated the importance of choosing 
parameters (especially carrier density) appropriate to the materials 
of interest by exhibiting the incorrect results obtained by the use of
wrongly chosen parameters. We have shown that the crucial quantity is 
the electron kinetic energy and have used our results to estimate the
kinetic energy of several manganite systems. Our results also provide 
insight into the crucial question of which portions of the spectrum 
pertain to the low energy electronic degrees of freedom.

There are several directions for future work. One is to combine the 
dynamical mean field method with a realistic band structure, to obtain 
a treatment of the frequency dependence of $\sigma$. Another is to
employ the methods presented here to models where double-exchange is 
combined with other interactions. If this were carried through, it 
seems likely that the combination of $T_c$ and the changes in optical 
spectral weight could be analysed to provide detailed knowledge of the
strength, energy scale, and nature of any additional couplings.

\section{Acknowledgements}

We thank H.D. Drew, B.G Kotliar and H. Monien for helpful discussions
and the University of Maryland MRSEC and NSF-DMR-9705482 for support.

\end{document}